# Mathematical Model for Detection of Leakage in Domestic Water Supply Systems by Reading Consumption from an Analogue Water Meter

Gal Oren and Nerya Y. Stroh

**Abstract**—In this article we introduce the principles to detect leakage using a mathematical model based on machine learning and domestic water consumption monitoring in real time. The model uses data which is measured from a water meter, analyzes the water consumption, and uses two criteria simultaneously: deviation from the average consumption, and comparison of steady water consumptions over a period of time. Simulation of the model on a regular household consumer was implemented on Antileaks – device that we have built that designed to transfer consumption information from an analogue water meter to a digital form in real time.

## I. Introduction

Water loss in supply systems is one of the most difficult problems of the water economy, which has numerous reasons [1]. Leakage reduces pressure in the supply system, while raising pressure to make up for such losses involves increased energy consumption, and has adverse environmental impacts. The financial damage caused to property should be taken into account as well [2].

The model presented here is monitoring the water flow, and determines if there is a leakage or not. The model is based on a statistical data analysis of the real time measurement of the flow [3], in order to give a simple and applicable engineering solution to domestic water supply systems, which can be supported by an analogue water meter, and not a digital one, as opposed to the rest of the market of leakage detectors.

To simplify the explanation we will focus on each criterion separately, even though the model requires all the criteria together.

## II. Non-existence of Leakage

An important and simple criterion to test the non-existence of leakage is characterized by a certain moment $t$ in which the level of consumption, $Cons(t)$, is zero (or pseudo-zero, since there may be a slight deviation in analog water meter measurements). Practically, it is defined as a detection of a very short time slot with no consumption.

Manuscript received January 10, 2013. This work was supported in part by the Jerusalem College of Technology.
Gal Oren, Author, is with the Jerusalem College of Technology and the Open University of Israel (e-mail: galoren.com@gmail.com).
Nerya Y. Stroh, Author, is with the Jerusalem College of Technology and the Open University of Israel (e-mail: nerya.s@gmail.com).

## III. Leakage detection based on consumption averages

### A. General description

To detect a leakage with this criterion, it is necessary to collect data about the average consumption ($avg$), the length of the time period in which the consumption was measured ($T$), and the maximum deviation level ($MD_T$) for this time period, which is the consumption threshold allowed to exceed the average consumption.

The time periods for the model pre-defined (such as: 15 minutes, 30 minutes, 2 hours, and 24 hours). *Potential leakage* is defined as the first deviation for a specific time period $T_1$ above the maximum deviation level. Thereafter, should another deviation over a longer time period $T_2$ be detected, the criterion will detect a leakage:

$$MD_{T1} < Consumption(T_1) \text{ AND} \\ MD_{T2} < Consumption(T_2) \quad (1)$$

### B. Distribution of consumption over time

Time periods shorter than or equal to a full day will be referred to as *short time periods*. In these cases, there might be radical changes in consumption levels relative to time (e.g. 5 minutes with certain consumption followed by 100 minutes with no consumption at all), that will cause large variance and significant lower averages, hence, the distribution of consumption will not be normal [4]. Therefore, these *short time periods* are defined not only by their length, but also by their *starting time* and *ending time* (e.g. a single average for the 30 minutes from 06:00 to 06:30 in each day). The distribution of the consumption in these time periods looks much more like a normal distribution.

Over *long time periods* (longer from or equal to full day), the changes that might be in the domestic consumption are not radical, and these cases will not be discussed here.

### C. Deviation level

The value of the $MD_T$ is determined using *statistical hypotheses*. Several case studies of a standard household's daily consumption were taken into account to establish the primary hypothesis, from Israel [5], United States [6], and Germany [7]. Using these cases, an estimation of daily domestic water usage (in Liters) relative to family size can be simulated with the following equation:

$$FC_t(S) = p \cdot S + 140 \quad (2)$$

Where $S$ is a family size, $FC_t(S)$ is the daily consumption of the family measured over time period $t$, and the coefficient $p$ is a normally distributed random variable, $p \sim N(120,20)$. This equation means that there is an increase of roughly 120 Liters for each person added to the family.

The steady-state leakage is simulated using a simple volumetric flow rate:

$$Q = \iint_A V \cdot dA \qquad (3)$$

Where $Q$ measure the water flow that leaks from the pipe in liters per seconds ($L \cdot s^{-1}$), $A$ is the cross-sectional area of the pipe orifice ($m^2$), and $V$ is the velocity field ($m \cdot s^{-1}$) in time $t$.

An accurate equation that is based on the pressure level and other parameters of the pipe can be developed [8], but there is no need for this in the current simulation.

The length of the *primary learning period* is two weeks, during which detection of leakage will be carried out using other criteria (see section IV). A numerical maximum deviation is set for situations where the consumption average is pseudo-0 even after the primary learning period.

The confidence level $(1 - \alpha)$ is 0.99 or 0.95. Where $n$ is the sample size, $\bar{X}$ is the mean estimator and the $\hat{S}$ standard deviation estimator is.

Leakage detection means specifically a deviation above the averages, resulting in the null hypothesis being right-tailed. According to the confidence interval for the mean $\mu$ of a normal population, when the variance is unknown, the upper critical value is:

$$K = \bar{X} + t_{1-\alpha}^{n-1} \cdot \frac{\hat{S}}{\sqrt{n}} \qquad (4)$$

The *t-scores* are pre-calculated using Student's t-distribution. If $n > 30$, we can substitute the $\hat{s}^2$ for $\sigma^2$ and use Z-test. The general formula of the $MD_T$ for time period $t$ with length $T$ is given by:

$$MD_T = \begin{cases} c(T), & K \leq 0.3 \\ a(T)K + b(T)\hat{S}, & otherwise \end{cases} \qquad (5)$$

Where the coefficients $a, b, c \geq 0$ tune the model behavior. Thus coefficients are set for each time period $t$ based on the studied cases data, the simulations, and the consumption patterns.

*D. Consumption patterns*

Three major consumption patterns in the domestic consumption were defined in the model: low consumption, stable non-low, and mutable non-low.
1) Low consumption pattern: the water consumption is low or non-existent. This pattern occurs when the family is asleep, and is also known as a *night flow* [9].
2) Stable pattern: the family has a scheduled and predictable consumption. This pattern occurs in the morning while most of the family is awake and home, and in the evening before they go to sleep.
3) Mutable pattern: most of the family is outside the house. This pattern occurs during the remaining daylight hours.

The problem of matching each consumption pattern to time period $t$ (with length $T > 300\ minutes$) is a machine learning classification problem.

The group of samples $X(t)$ that begins at $SP = T_0$, and ends at $EP = T_0 + T$ with *interval time* ($IT \in \mathbb{Z}$) between the samples, is given by:

$$X(t) = \{x_i = Cons\ (T_0 + IT \cdot i)\ |\ i \in \mathbb{N},\ IT \cdot i \leq T\} \qquad (6)$$

Where $Cons(t)$ is the measured consumption at $t$. For group of samples $Y = \{X_d(t), D_1 \leq d \leq D_2\}$ at time period $t$, between the days $D_1$ to $D_2$, the classifying function $\phi_1(t, Y(t))$ determines whether a low consumption pattern was detected:

$$\phi_1(t, Y(t)) = \begin{cases} Low, & \frac{6}{7}(D_2 - D_1) \leq \sum_{i=D_1}^{D_2} Low(t, X_i(t)) \\ NonLow, & otherwise \end{cases} \qquad (7)$$

Where $Low(t, X(t))$ checks whether a group of samples $X$ over period time $t$ is not above 15 Liters:

$$Low(t, X(t)) = \begin{cases} 1, & Cons_t(X) \leq 15 \\ 0, & otherwise \end{cases} \qquad (8)$$

If a non-low consumption pattern was detected, for the consumptions vector $\mathbf{Z} = Z(t) = \{z_i = \|X_i(t)\|, D_1 \leq i \leq D_2\}$, at a time period $t$ between the days $D_1$ to $D_2$, the function $\phi_2(t, Z(t))$ classifies the consumption state:

$$\phi_2(t, Z(t)) = \begin{cases} Stable, & S_Z < 20/3 \\ Mutable, & otherwise \end{cases} \qquad (9)$$

Where $\bar{X}_Z$ and $\hat{S}_Z$ are the mean estimator and the standard deviation estimator of $\mathbf{Z}$, respectively. This implies that the classifier $\phi_2(t, Z(t))$ checks whether the samples tend to gather around the mean.

As mentioned, each pattern determines the value of the coefficients $a, b$ of $MD_T$ for time period $t$ with length $T$.

Let $\mathbf{STP} = (T_1, T_2, .., T_{RL})$ be the vector of the short time periods defined in the model, where $RL \in \mathbb{N}$ is the resolution level of $\mathbf{STP}$. The coordinates of the matrix $\mathbf{A} = (a1(T_n), a2(T_n), a3(T_n))$ represent the values of the $a$ coefficient at time period $t$ with length $T_n \in \mathbf{STP}$, for given consumption pattern $CP$ (low, stale, and mutable respectively). Thus values are determined manually.

For $\mathbf{STP1} = (15, 30, 60, 120, 300, 480, 720)$, they are shown at Fig. 1. Similarly, the values of $\mathbf{B} = (b1(T_n), b2(T_n), b3(T_n))$ are shown at Fig. 2.

The values of $\mathbf{C} = (c(T_n))$ at time period $t$ with length $T_n, n \leq RL$, are similar for each consumption pattern, and are given by:

$$c(T_n) = 20(1 + n) \qquad (10)$$

If the time period $t$ spans over $M$ consumption patterns, with length $T = \sum_{m=1}^{M} T_m$, the $MD_T$ formula is:

$$MD_T = \sum_{m=1}^{M} (T_m/T)\, MD_{T_m} \qquad (11)$$

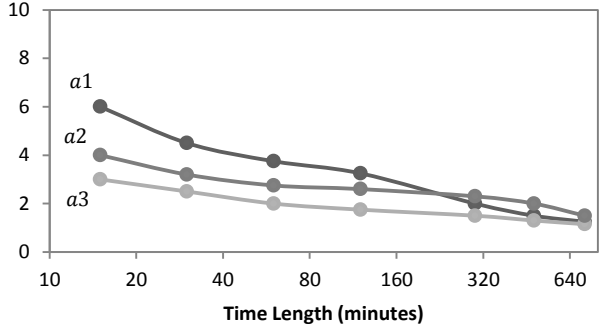

Fig. 1. The values of $a$ at time period $t$ with length $T \in \mathbf{STP1}$, for given consumption pattern $CP$ (1=low, 2=stable, 3=mutable)

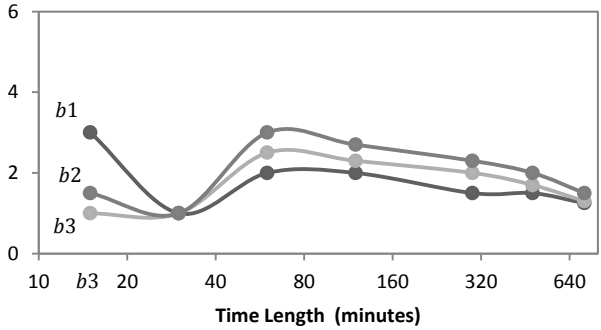

Fig. 2. The values of $b$ at time period $t$ with length $T \in \mathbf{STP1}$, for given consumption pattern $CP$ (1=low, 2=stable, 3=mutable)

### E. Result of Simulation

The following graphs represent the simulation results of this criterion [10]. In Fig. 3. for $T_1 = 15$ minutes, two potential leakages were detected..

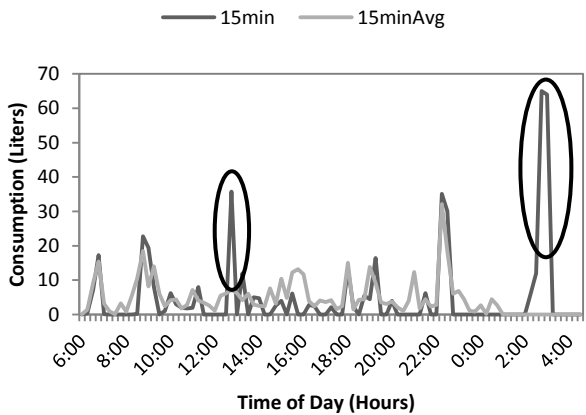

Fig. 3. The general consumption averages and the specific consumption for $T_1 = 15$ minutes on a simulated leakage day.

Fig. 4 shows the same measures for $T_2 = 30$ minutes, where the first alert was interpreted as a regular usage, and the second alert was interpreted as a real leakage.

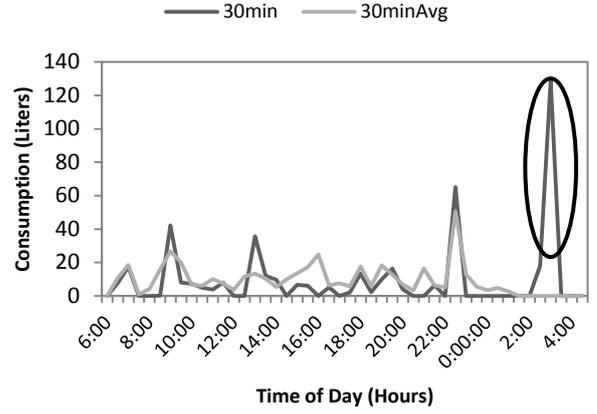

Fig. 4. The general consumption averages and the specific consumption for time period $T_2 = 30$ minutes on a simulated leakage day.

## IV. LEAKAGE DETECTION BASED ON STEADY WATER CONSUMPTION OVER TIME

### A. General description

*Steady water consumption* is referred to a time period *TP* (that begins at $T_0$) with no interruption of the consumption. The *interval time* (*IT*) between the samples needs to be short. By this criterion a leakage can be detected while the system is still in the primary learning period.

In the consumer household, the chances for a long steady consumption are very low (even heavy-usage devices such as a washing machine consumed water in pulses).

Practically, there might be a real household consumption in parallel to the undesired leaking water consumption from the system. This case will be referred to *steady water consumption of an AC wave riding on a DC wave*. A simple (yet effective) way to detect a leakage in such case is to use a filter that checks if the majority of the water consumption samples (more than 50%) revolve around the *median* (*med*) – namely, the DC wave – with a *slight deviation* (*SD*). It can be said that a leakage is detected for a group of samples $X$ using (6) if:

$$\begin{aligned} & x_i > 0.2 \text{ AND} \\ & 0.5 < \frac{1}{n}\sum_{i=0}^{n}(x_i \in [Med(X) \pm Med(X) \cdot SD]) \end{aligned} \qquad (12)$$

### B. Result of simulation

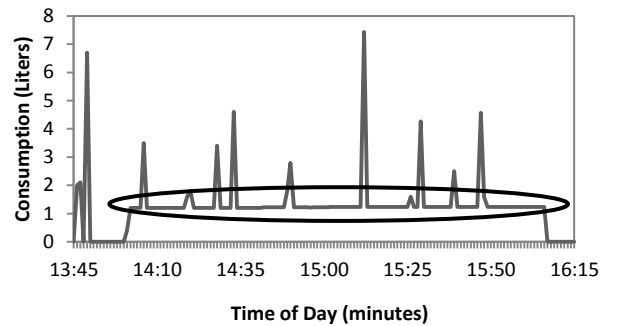

Fig. 5. Consumption during 13:45-16:45.

Fig. 5 represents the simulation result of this criterion [10]. Steady water consumption pattern was detected during 14:02-16:02, and was interpreted as a real leakage.

## V. Handling false leakage detection

The false alert (false negative) is handled by a simple human-computer decision making. The coefficient $r$ is measuring the model reliability which is:

$$r = \begin{cases} (AN - FN)/LN, & AN > 0 \\ 1, & AN = 0 \end{cases} \quad (13)$$

Where $AN$ is the amount of alerts, $FN$ is the amount of false alerts, and $LN$ is the amount of real leakages in the supply system.

Tuning of the $MD_T$ to $TMD_T$ (Tuned $MD_T$) performed after the primary learning period, for each reported alert:

$$TMD_T = \begin{cases} \frac{1}{2}(AN + 1.1)MD_T, & r = 0, AN > 0 \\ \frac{1}{r}MD_T, & r > 0 \end{cases} \quad (14)$$

## VI. Edge cases of consumption patterns

A fire alarm is a feature of a *building management system (BMS)* that is common in large buildings [11]. In fire emergency situations, an incensement in water consumption is predicated. This edge case can be detected as an unusual consumption pattern, and interpreted as a leakage. While implementing the model in a leakage detection device, it is necessary to synchronize the detection device with the fire alarm system.

Few other issues can impact measured consumption, such as air pockets flowing through water meters that cause small "fake" consumptions [12]. These air pockets are filtered from the model using the slight deviation parameter in the steady water consumption criterion.

## Acknowledgment

We would like to thank Eng. David Gelman and Dr. Emanuel Resin from the Jerusalem College of Technology, and Prof. Gedeon Dagan from Tel Aviv University for scientific advice.

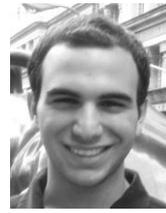

**Gal Oren** was born in Israel on July 7th 1991. Department of Computer Science, B.Sc., Jerusalem College of Technology (2006-2010); Department of Computer Science, M.Sc., Open University of Israel (2011-Present). He is the co-inventor (with Nerya Y. Stroh) of Antileaks – a Device for Detection and Discontinuation of Leakages in Water Supply Systems.

Mr. Oren was awarded the following prizes: Zalman Shazar Prize (2009), Commendation from the President of Israel for an original scientific study (2010), Stockholm Junior Water Prize in Israel, Tel Aviv University (2010), Honor Diploma from Her Royal Highness Crown Princess Victoria of Sweden, Stockholm Junior Water Prize, World Water Week, Stockholm, Sweden (2010), First Prize in the Technology field at the National Young Scientist and Engineers Competition in Israel, Hebrew University of Jerusalem (2011), Intel Prize at the European Union Contest for Young Scientists, Helsinki, Finland (2011), Third Grand Award in Environmental Management category and the Third Special Award in Sustainable Water Management category at the Intel International Science and Engineering Fair, Pittsburgh, Pennsylvania, the United States (2012), and the Encouraging Artistic Creativity Among Scientists Prize, Weizmann Institute of Science (2013).

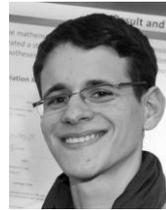

**Nerya Y. Stroh** was born in Israel on September 5th 1991. Department of Computer Science, B.Sc., Jerusalem College of Technology (2006-2010); Department of Computer Science, M.Sc., Open University of Israel (2011-Present). He is the co-inventor (with Gal Oren) of Antileaks – a Device for Detection and Discontinuation of Leakages in Water Supply Systems.

Mr. Stroh was awarded the following prizes: Stockholm Junior Water Prize in Israel, Tel Aviv University (2010), Honor Diploma from Her Royal Highness Crown Princess Victoria of Sweden, Stockholm Junior Water Prize, World Water Week, Stockholm, Sweden (2010), First Prize in the Technology field at the National Young Scientist and Engineers Competition in Israel, Hebrew University of Jerusalem (2011), Intel Prize at the European Union Contest for Young Scientists, Helsinki, Finland (2011), Third Grand Award in Environmental Management category and the Third Special Award in Sustainable Water Management category at the Intel International Science and Engineering Fair, Pittsburgh, Pennsylvania, the United States (2012).